\documentstyle[twocolumn,ucsd,epsf]{article}

\def\roughly#1{\raise.3ex
    \hbox{$#1$\kern-.75em\lower1ex\hbox{$\sim$}}}
\def\be{\begin{equation}}
\def\ee{\end{equation}}
\def\bea{\begin{eqnarray}}
\def\eea{\end{eqnarray}}

\newcommand{\square}{\kern1pt\vbox{\hrule height 0.6pt\hbox{\vrule width
0.6pt\hskip 3pt
   \vbox{\vskip 6pt}\hskip 3pt\vrule width 0.6pt}\hrule height 0.6pt}\kern1pt}
\begin{document}

\renewcommand{\thepage}{\roman{page}}
\renewcommand{\thebean}{\roman{bean}}
\setcounter{page}{-1}
\setcounter{bean}{0}

\begin{center}
March 1993\hfill   UCSD-PTH 93-08

\vskip .6in

\renewcommand{\thefootnote}{\fnsymbol{footnote}}
\setcounter{footnote}{0}

{\large\bf
Strongly Interacting Higgs Sector in the Minimal Standard
Model\,?\,}\footnote{ \noindent This work was supported by the
U.S. Department of Energy under Grant\newline \indent\indent
DE-FG03-91ER40546.}

\renewcommand{\thefootnote}{\alph{footnote}}
\setcounter{footnote}{0}

\vskip .4in

Karl Jansen,\footnote{E-mail: jansen@higgs.ucsd.edu}
Julius Kuti,\footnote{E-mail: kuti@sdphjk.ucsd.edu}
Chuan Liu\footnote{E-mail: chuan@higgs.ucsd.edu}

\vskip 0.2in

{\em Department of Physics 0319\\
     University of California at San Diego \\
        9500 Gilman Drive\\
        La Jolla, CA 92093-0319 USA}

\vskip 2cm

{\em Submitted to Physics Letters B}

\vskip 3cm

{\em Part Two of Extended UCSD-PTH 92-40}

\end{center}

\newpage

\mbox{}

\vskip 0.7in

\begin{center}
{\bf Abstract}
\end{center}

\vskip 0.2in

\begin{small}
\begin{quotation}
The triviality Higgs mass bound is studied with a higher derivative regulator
in the spontaneously broken phase of the four dimensional O(4) symmetric scalar
field theory with quartic self-interaction.
The phase diagram
of the O(4) model is determined in a Monte Carlo simulation which
interpolates between the hypercubic lattice regulator and the higher
derivative regulator in continuum space-time. The same method can be used
to calculate the
Higgs mass bound in continuum space-time.
In a large-N analysis, when compared with
a hypercubic lattice, we find a relative increase in the triviality bound
of the higher derivative regulator
suggesting a strongly interacting Higgs
sector in the TeV region with negligible dependence on regulator parameters.
When the higher derivative regulator mass is brought close to the Higgs mass
the
model requires a  more elaborate analysis of complex ghost states in scattering
amplitudes.
\end{quotation}
\end{small}

\newpage

%\noindent PACS numbers: 04.60.+n,11.30.Pb,12.10.-g.

%\mbox{ }

\renewcommand{\thepage}{\arabic{page}}
\renewcommand{\thebean}{\arabic{bean}}
\setcounter{page}{0}
\setcounter{bean}{0}

\renewcommand{\thefootnote}{\arabic{footnote}}
\setcounter{footnote}{0}

%THIS IS PAGE 1 (INSERT TEXT OF REPORT HERE)

\section{Introduction}

We develop a systematic method to investigate
in continuum space-time the possibility of a strongly interacting Higgs sector
in the minimal Standard Model with a heavy Higgs particle in the TeV
region. This question, which is directly relevant for the Higgs physics program
of the Superconducting Supercollider Laboratory, has never  been answered in
a reliable fashion, although speculations and phenomenological approximations
are well-known. We report here some results on our non-perturbative
investigation
of the problem.

In earlier lattice studies
the triviality upper bound for the Higgs mass
was found at 640 GeV
under some well defined set of conditions \cite{KLS,LW,HJJLNY}  with a
lattice
momentum cut-off at 4 TeV. One can show from the equivalence theorem \cite{LQT}
that lattice artifacts with the 4 TeV cut-off remain hidden to a few percent
accuracy below 1.3 TeV  center of mass energy (the physics reach of the SSC) in
the experimentally relevant $W_LW_L$ cross sections,
if the Higgs mass is kept below the upper bound. There has been great
concern that this finding was an artifact of the lattice regulator which breaks
Euclidean invariance.
In fact, the first significant increase of the Higgs mass bound was reported
\cite{GKNZ}
within the Symanzik improvement program on a hypercubic lattice structure
\cite{SYM}. Similar results on different lattice structures, with
higher dimensional lattice operators in the interaction term, were also
reported
\cite{HNV}.

In this work we replace the lattice regulator with a higher
derivative kinetic term in the Higgs Lagrangian which, according to
conventional thinking, acts as a Pauli-Villars regulator preserving all the
relevant symmetries of the theory. We will investigate the regulator dependence
of the theory without lattice artifacts.
The phase diagram
of the O(4) model will be determined in a Monte Carlo simulation which
interpolates between the hypercubic lattice regulator and the higher
derivative regulator in continuum space-time. The same method can be used
to calculate the triviality
Higgs mass bound in continuum space-time.
In a large-N analysis\footnote{Variants of the large-N analysis in the study
of the O(N) model were developed earlier\cite{HNV,EW,KLS2}.}, when compared
with
a hypercubic lattice, we find a relative increase in the triviality bound
of the higher derivative regulator
suggesting a strongly interacting Higgs
sector in the continuum with negligible dependence on the regulator mass
parameter.

Consider the euclidean action of the
O(4) symmetric
scalar field theory in four dimensions with a higher derivative term
in the kinetic energy,
\begin{equation}
%\lefteqn{
S_E  = \int d^4x \Bigl[ - {1\over2} ~
\vec\phi~(\square + {1\over M^4}\square^3)~\vec\phi
%}
% \mbox{}~~~~~~~~~~
- {1\over2} m_0^2~\vec\phi\cdot\vec\phi + {\lambda_0}
(\vec\phi\cdot\vec\phi)^2 ~ \Bigr] ~ ,
\label{eq:SEO4}
\end{equation}
where the scalar field $\vec\phi$ has 4 components and $\square$ is the
euclidean Laplace operator; the bare parameters
$m_0^2$, $\lambda_0$ are the same as in the ordinary O(4) theory.
The higher
derivative term $M^{-4}\square^3$ acts as a Pauli-Villars regulator
with mass parameter $M$. It represents a minimal modification of the
continuum model,
if we want to render the field theory and its euclidean path integral,
\begin{equation}
         Z = \int [d\vec\phi]~{\rm exp} ~ \Bigl\{ -S_E[\vec\phi] \Bigr\}\; ,
\label{eq:ZE1}
\end{equation}
finite.

To understand the physics of higher derivative regulator effects
we had to derive the euclidean path integral from the Hamiltonian formulation
in the Hilbert space of quantum states. We also had to understand
the spectrum of the Hamiltonian, the role of ghost
states in scattering processes, and the  closely related issues of
unitarity and microscopic acausality effects. Results along these lines were
reported in previous publications \cite{JKL1,JKL2}.

\section{Phase Diagram and Pauli-Villars Limit}

For non-perturbative computer investigation of the higher derivative
Lagrangian we introduce a hypercubic lattice
structure. The
lattice spacing $a$ defines a new short distance scale in the theory with the
associated lattice momentum cut-off at $\Lambda = \pi/a$. We will have to
work towards the large $\Lambda/M$ limit of the continuum theory in order to
eliminate finite lattice effects from the already regulated and finite theory.
The presence of the lattice parameter $a$ allows us to
smoothly interpolate between the lattice regulated theory and the finite
Pauli-Villars theory in the continuum. In computer simulations the lattice
spacing $a$ is set to one for convenience, and $m_0^2$,
$M^2$ are measured in lattice units.

After rescaling the continuum
fields by $\vec\phi =\vec\varphi\sqrt{2\kappa}$ the euclidean lattice action
can be written as
\begin{equation}
%\lefteqn{
S_E  = \sum_x \Bigl[ - \kappa ~
\vec\varphi~(\square + {1\over M^4}\square^3)~\vec\varphi
+ (1-8\kappa)\vec\varphi\cdot\vec\varphi + {\lambda}
(\vec\varphi\cdot\vec\varphi-1)^2 ~ \Bigr] ~ ,
\label{eq:SEO4L}
\end{equation}
where
the continuum operator $\square$ is replaced  by the equivalent hypercubic
lattice operator
$\square=\sum_{\hat\mu}
(\delta_{x,x+\hat\mu}+\delta_{x,x-\hat\mu}-2\delta_{x,x})$.
The lattice points are labeled by the variable $x$,
and $\hat\mu$ designates unit vectors along the four independent directions.
The mass parameter and coupling constant are related to the hopping parameter
$\kappa$ and the lattice coupling $\lambda$ by
$m_0^2=(1-8\kappa-2\lambda)/\kappa$, $\lambda_0=\lambda/4\kappa^2$. In
the limit of infinite $\lambda$ the radial mode of the scalar field becomes
frozen at $|\vec\varphi_x| =1$ in our normalization.
This is the limit of the nonlinear $\sigma$-model where we
expect to find the maximal upper bound for the Higgs boson mass.

The lattice model defined by Eq.~\ref{eq:SEO4L} can be studied
non-perturbatively in computer simulations.
The standard and popular stochastic algorithms which are based on
local updating schemes are ineffective when applied to this particular
lattice model, exhibiting severe
problems of critical slowing down.
The very long autocorrelation times became particularly prohibitive
for small values of the Pauli-Villars mass parameter, in the range $M < 1$.
In this regime the $M^{-4}p^6$ term of the kinetic energy in the higher
derivative lattice Lagrangian leads to a dramatic broadening of the
eigenfrequency spectrum
of the Fourier modes which is the primary source of very long
autocorrelations.
We developed a version of the Hybrid Monte Carlo Algorithm (HMC)
where the eigenmodes were accelerated
by Fast Fourier Transformation (FFT) techniques. This new algorithm solved the
problem of critical slowing down very effectively. The HMC algorithm
with FFT techniques allowed us to study the lattice model of Eq.~\ref{eq:SEO4L}
at the physically important values of the parameters.

The lattice model defined by the euclidean partition function has
two phases, as expected.
The phase diagram is shown in Fig.~1, in the $\lambda=\infty$ limit.
The phase transition line is plotted as the
dotted line from a large-N calculation and
the full circles represent our simulation results. The
large-N approximation also predicts that the phase transition is of second
order
along the whole transition line, in
agreement with the simulation results.
In the symmetric phase
we find the original massive particle (with four components in the
intrinsic O(4) space), and also a complex ghost pair
with intrinsic
O(4) symmetry whose mass scale is set by the Pauli-Villars
mass parameter $M$.
In the broken phase a Higgs
particle was found with mass $m_H$, and three massless Goldstone excitations
with residual O(3) symmetry. In addition,
the 4-component heavy ghost particle and its complex conjugate partner
also appear in the spectrum of the
broken phase.

The continuum limit, which eliminates the underlying
lattice structure, is equivalent to the tuning of $\kappa$ to its critical
value for fixed bare coupling, so that the physical Higgs mass $m_H \to 0$
in lattice units. At the same time the physical (renormalized) mass $M_R$ of
the
regulator particle also decreases towards $M_R \to 0$ in lattice units,
but the ratio $m_H/M_R$ has to be kept
fixed by the appropriate adjustment of $M$. In this limit the
higher derivative continuum theory is obtained with finite mass scale for the
ghost states in Higgs mass units.
The solid line in Fig.~1 displays the fixed $M/m_H=10$ ratio in a large-N
calculation.
Moving on the line of constant $M/m_H$ towards the $aM \to 0$ limit is
indicated
by arrows as we are approaching
the higher derivative continuum theory.
At fixed lattice spacing $a$, the $aM \to \infty$ limit will produce the
ordinary O(4) field
theory on a hypercubic lattice. Approaching the dotted critical line for
any fixed
$M$ corresponds to the limit of trivial free field theory.
\section{The Higgs Mass Bound and Large-N}

The Higgs mass bound in
the simple O(4) model on a hypercubic lattice structure was determined earlier
\cite{KLS,LW,HJJLNY}. The calculation had a few important ingredients.
To obtain nonvanishing  $m_H/v$ values we had to move away from the
critical line of the phase diagram where the ratio vanishes ($v$ denotes the
vacuum expectation value of the Higgs field). The ratio $m_H/v$ grows as a
function of decreasing
$\Lambda/m_H$, when
moving away from the critical line. An approximate upper bound on $m_H/v$
exists in the region where  $\Lambda/m_H$ is
not large enough to keep lattice cut-off effects reasonably small in physical
quantities.
For fixed $\Lambda/m_H$ the largest $m_H/v$ ratio is obtained at infinite bare
Higgs coupling. When $\Lambda/m_H = 2\pi$ is chosen, one finds the ratio
$m_H/v = 2.6$ at $\lambda_0 = \infty$ which leads to the 640 GeV Higgs mass
bound. It corresponds to a rather small renormalized quartic
coupling constant, excluding a strongly interacting Higgs sector in the
simple O(4) model with hypercubic lattice regulator.

The reason one should not push the lattice cut-off much lower than
$\Lambda/m_H = 2\pi$ is as follows.
In physical cross sections lattice cut-off effects
(breakdown of Euclidean invariance) begin to appear when $\Lambda/m_H$
gets smaller.
A somewhat arbitray but useful measure of
lattice cut-off effects is the ratio $R$ of the perturbative continuum cross
section for elastic Goldstone particle scattering at $90^\circ$ and
the equivalent regulator dependent cross section at finite cut-off.
Deviations from $R=1$ indicate the size of cut-off dependence \cite{LW}.
For $\Lambda/m_H \geq 2\pi$, and center-of mass energies $W\le\Lambda/\pi$, the
cut-off effects in $R$ do not exceed
the few percent level \cite{LW}.
In the region $\Lambda/\pi\leq W\leq 2\Lambda/\pi$
there is a rapid growth in $R$, and for $W\geq2\Lambda/\pi$ the cut-off effects
become large, as
expected when one gets close to the lattice momentum cut-off.

As a test of the above scenario on lattice regulator dependence, the Higgs
mass bound can be estimated in the large-N approximation as we interpolate
between the simple lattice O(N) model and the continuum O(N) model with
Pauli-Villars regulator. The mass of the Higgs particle
is given by
\begin{equation}
\frac{m_H}{M} = C(aM,\lambda_0)\cdot {\rm exp} \Bigl\{-\frac{16\pi^2
v^2}{N\cdot m_H^2} \Bigr\} ~,
\label{eq:C_PV}
\end{equation}
where the amplitude $C(aM,\lambda_0)$ is calculable from a Goldstone
loop diagram.
At fixed $M/m_H$,
$C_{PV}(\lambda_0) = \lim_{\, aM\to 0} C(aM,\lambda_0)$
is the higher derivative (Pauli-Villars) regulator limit of the amplitude
in continuum
space-time. At fixed $a$, $C_{LAT}(\lambda_0) = \lim_{\, aM\to \infty}
[aM \cdot C(aM,\lambda_0)]$
corresponds to
the original hypercubic lattice theory without higher derivative
regulator term (the reason for the prefactor $aM$ in front of the
amplitude $C(aM,\lambda_0)$ is the replacement of $m_H/M$ by $a\cdot m_H$
on the left side of Eq.~\ref{eq:C_PV}  in the $M\to \infty$ limit).
Numerically both amplitudes are the largest in the $\lambda_0 \to
\infty$ limit where we find $C_{PV}= e^{1/4}$ and $C_{LAT}=e^{2.896}$,
for $N=4$.

Using Eq.~\ref{eq:C_PV} and the numerical value of $C_{LAT}$ we plot $m_H/v$
as the dotted line of Fig.~2a against the lattice correlation length
$\xi_L=(a\cdot m_H)^{-1}$ of the simple O(4) model
at $\lambda_0 =\infty$. The solid line is the plot
of $m_H/v$ against $\xi_{PV}=M/m_H$ in the higher derivative continuum theory,
with Pauli-Villars regulator mass $M$ and $C_{PV}$ in Eq.~\ref{eq:C_PV}.
The two independent curves are plotted together by
using a joint numerical $\xi$-axis for the convenience of comparison.
Since $C_{PV}$ is significantly smaller than $C_{LAT}$, the solid $PV$ curve
runs
above the dotted lattice curve.

As it was found earlier,
in the $aM \to \infty$ limit, $m_H\cdot a = 0.5$ corresponds to
few percent lattice effects
in the physical cross section
of elastic $W_LW_L$ scattering at  center of mass energies $W \leq 2m_H$
\cite{LW}.
In Fig.~2a arrow points to the dotted curve at this Higgs correlation length
representing the upper bound of the simple O(4) model on the hypercubic lattice
in the large-N approximation.
At fixed $M/m_H = 4$,  in the $a\cdot M \to 0$ Pauli-Villars limit, we find a
comparable few percent deviation from
$R=1$
(ghost effects) in the $W_LW_L$ cross section
of the higher derivative continuum theory
for $W \leq 2 m_H$, as shown in
Fig.~2b.
In the figure we show the ratio $R$ as a function of the center of mass
energy $W$ when evaluated in the Pauli-Villars
regulated theory in the continuum (dotted line), and the lattice model at
$a\cdot M = 0.8$ (solid line).
In Fig.~2a arrow points to the solid curve at $\xi_{PV} = 4$
representing the upper bound of the higher derivative continuum O(4) model
in the large-N approximation.
In fact, the ratio $M/m_H=4$ is a rather conservative choice since $R$ deviates
from 1  on the few percent level even at the much lower value
of $M/m_H \approx 3$ which would imply a larger Higgs mass bound.

A change in the Higgs mass bound  can be found by comparing the
ratio  $m_H/v$ in the two different schemes,
\begin{equation}
{(m_H/v)_{PV}\over (m_H/v)_{LAT}} = \sqrt{ {{\rm ln}\, (am_H)_{LAT} + {\rm
ln}\,
C_{LAT}(\lambda_0) \over {\rm ln}\, (M/m_H)_{PV} + {\rm ln}\,
C_{PV}(\lambda_0) }} ~~.
\label{eq:RATIO}
\end{equation}
Using the large-N numbers for $C_{LAT}$ and
$C_{PV}$ at $\lambda_0 = \infty$, we find from Fig.~3a and
Eq.~\ref{eq:RATIO} that the Higgs mass bound increases from $m_H  \approx 3v$
to
$m_H \approx 5v$ as we interpolate from the hypercubic lattice action
to the continuum
higher derivative regulator.
%Choosing a value for $M=1$ would mean
%that one can work at a lattice correlation lenght of around $\xi_L =3$ to
%%obtain
%a realiable estimate of the upper bound from lattice simulations.

In a related large-N analysis \cite{HNV} the $C_{LAT}$ and $C_{PV}$
coefficients
were also calculated.
However, the large relative increase of the Higgs mass bound in the higher
derivative continuum model with respect to the simple lattice model
was not pointed out.
We have to emphasize at this point that it is the {\em relative}
increase of the Higgs mass bound as described by Eq.~\ref{eq:RATIO} which is
the important hint from the large-N approximation. As it was pointed out
earlier \cite{KLS2},
$N=4$ is too far from $N=\infty$ for any reliable absolute estimate of
the Higgs mass bound by using the large-N expansion. The main problem is
that the $N+8$ factor in the $\beta$-function and other perturbative formulae
creates a mismatch between perturbation theory and large-N for $N=4$.
This mismatch
disappears only for $N\gg 8$.
The difficulty is clearly illustrated by Fig.~7.6 of the second paper in
reference \cite{HNV} where
the  width of the Higgs resonance is plotted as a function
of the Higgs mass $m_H$.
The Higgs self-coupling is
perturbative in the mass region $m_H \leq 700~ GeV$
and, therefore, the width of the Higgs resonance from the
large-N analysis
should agree with the perturbative calculation. Instead, a large
discrepancy is exhibited between perturbation theory and the large-N expansion
which cannot produce reliable absolute results at $N=4$.

It is indicative, however, that the relative increase of the Higgs mass bound
is
remarkably stable. Eq.~\ref{eq:RATIO}
was derived from the large-N result of Eq.~\ref{eq:C_PV} where the finite
width of the Higgs resonance was neglected. The finite width corrections, and
$N+8$ effects in Hartree approximation, lead to a more complicated numerical
analysis which lowers the absolute positions of the curves in Fig.~2a, but
the relative increase of the upper bound in the higher derivative continuum
theory, when compared with the simple O(4) model on the lattice, remains
approximately as large as in the simplest large-N expansion. It is suggestive
that the large relative increase of the upper bound is a robust feature of the
higher derivative theory. In this case we expect the Higgs mass bound to be
driven into the TeV mass range
{\em without} noticeable ghost effects in the accessible energy range of
$W_LW_L$ scattering at the
Supercollider. Our first simulation results on the mass spectrum
are supporting this scenario, with the possibility of a strongly
interacting Higgs sector in the minimal Standard Model.

\section{Conclusions}

We believe that there are two complementary views on the higher derivative
Lagrangian of the Higgs model. According to conventional thinking, the higher
derivative kinetic term acts as an ordinary cut-off on a mass scale set by M.
In this letter, we studied the Higgs mass bound under the condition that
cut-off effects associated with the regulator remain hidden from experiments.
The possibility of a strongly interacting Higgs sector was raised in
the minimal Standard Model, with {\em two} independent parameters
$m_H$ and $v$. Scattering amplitudes were restricted to the energy range
below the mass of the conjugate ghost pair where dependence on the mass scale
$M$ is almost negligible.

In a more unconventional approach, the higher derivative Lagrangian is
analyzed
as a logically consistent and finite field theory \cite{JKL2} with calculable
scattering amplitudes at arbitrary energies. In this scenario a heavy ghost
particle with complex mass is added to the minimal Higgs sector with observable
physical consequences. A strongly interacting Higgs sector emerges naturally,
but the physical scattering processes depend on the new mass parameter $M$
in a nontrivial fashion.

The non-perturbative computer investigation of the higher derivative Lagrangian
required the introduction of an underlying hypercubic lattice structure.
It is important to work in the large $\Lambda /M$
limit where lattice effects are negligible compared with ghost effects, so that
the finite continuum theory is simulated. We hope to return to further
non-perturbative investigations of both viewpoints in the higher derivative
Higgs model.

\subsection*{Acknowledgements}

We thank Jeroen Vink and other members of the Particle Theory Group at U.C. San
Diego for useful conversations and comments. We are also grateful to Howard
Georgi and a referee who suggested improvements in the presentation.
This work was supported by the DOE under Grant DE-FG03-91ER40546.
The simulations were done at Livermore National Laboratory with
DOE support for supercomputer resources.

\newpage

\section*{Figure Captions}
\vskip 1cm
\begin{description}

\item[Fig.~1:]
The phase diagram of the interpolated Pauli-Villars lattice model
at infinite bare coupling is shown. The dotted
line is the analytic form of the critical line in the large-N approximation.
The data points of critical $\kappa$ values are simulation results in
the O(4) model.
The solid line displays the fixed $M/m_H=10$ ratio towards the $aM \to 0$
Pauli-Villars limit of the higher derivative continuum theory;
it is
stretched by a factor of 5 along the $\kappa$-axis
for better separation from the phase transition line.

\item[Fig.~2:]
(a) The ratio $m_H/v$ is plotted in the large-N approximation of the lattice
model and the higher derivative continumm model
at $\lambda_0 = \infty$ using a joint $\xi$-axis for the convenience of
comparison; $\xi_L = am_H$, and $\xi_{PV} = M/m_H$, repectively. Arrows mark
the
respective Higgs mass bounds.
(b) The ratio $R$ of the perturbative continuum cross
section for elastic Goldstone particle scattering at $90^\circ$ and
the equivalent regulator dependent cross section at finite cut-off is plotted.
The dotted line represents $R$ in the large-N calculation with
Pauli-Villars regulator in the continuum theory, and the solid line is the
same calculation for the
lattice model of the higher derivative theory. Deviations from $R=1$ indicate
the size of regulator effects.

\end{description}

\vfill
%\pagebreak

\begin{figure}[t]
\centerline{\epsfbox{08_fig1.ps}}
\end{figure}

\begin{figure}[t]
\centerline{\epsfbox{08_fig2.ps}}
\end{figure}

%\begin{figure}[t]
%\centerline{\epsfbox{fig2.ps}}
%\end{figure}

%\vskip 5in
%\special{postscriptfile fig1.ps}

%\pagebreak

%\special{postscriptfile fig2.ps}

\end{document}